\documentclass[10pt,aps,nofootinbib,
amsmath,amssymb,twocolumn,
preprintnumbers, 
showpacs,
raggedbottom, 
superscriptaddress,
prl,
floatfix]{revtex4-2}

\usepackage{cases}
\usepackage{pgf,tikz}
\usepackage{graphicx}
\usepackage{booktabs}

\usepackage{slashed} 

\usepackage{hyperref}
\usepackage[nameinlink]{cleveref}
\crefdefaultlabelformat{#2\textup{#1}#3}
\creflabelformat{equation}{#2\textup{#1}#3}

\newcommand{\lrp}[1]{\left(#1\right)} 
\newcommand{\lrb}[1]{\left[#1\right]} 

\newcommand{\tov}{\text{TOV}}

\newcommand{\ceft}{\chi\text{EFT}}

\newcommand{\eden}{\mathcal{E}}

\newcommand{\mchirp}{M_\mathrm{chirp}}
\newcommand{\msol}{\mathrm{M_\odot}}

\newcommand{\UCB}{Department of Physics, University of California Berkeley, Berkeley, CA 94720}

\begin{document}

\title{Dwarf (Twin) Neutron Stars I: Did GW170817 Involve One?}

\author{Dake Zhou}
\email{dkzhou@berkeley.edu}

\affiliation{Department of Physics, University of Washington, Seattle, WA 98195}
\affiliation{Institute for Nuclear Theory, University of Washington, Seattle, WA 98195}
\affiliation{\UCB}
\affiliation{Department of Physics and Astronomy, Northwestern University, Evanston, IL 60208}

\preprint{N3AS-23-027}

\date{\today}

\begin{abstract}

Dwarf neutron stars are stable twins of neutron stars but with a maximum mass less than that of neutron stars. 
Their existence brings into concordance the seemingly conflicting information on the size of neutron stars inferred from gravitational waves from GW170817, from the NICER mission, 
and from the PREX-II experiment.
Their distinctive characteristics lead to rich and falsifiable predictions that are expected to be tested in the near future.
If corroborated, the existence of dwarf neutron stars would substantially improve our understanding of the QCD phase diagram and offer valuable insights into the dark sector.

\end{abstract}

\maketitle

The zero-temperature QCD phase diagram remains a mystery.
While it is clear that nuclear matter and quark matter occupies the low- and high-density corners,
we do not know if the liberation of quarks at finite densities involves latent heat
and if additional exotic phases exist in the intermediate region~\cite{Son:1998uk,Rajagopal:2000wf,Alford:2007xm}.
If there is a sizable first-order phase transition that occurs below baryon chemical potentials about $\mu_B\lesssim 2.2$ GeV, 
a third-generation of compact stars arise.

Traditionally, this possibility is severely constrained by the existence of two-solar-mass neutron stars~\cite{Demorest:2010bx,Antoniadis:2013pzd,Romani:2021xmb,NANOGrav:2019jur,Fonseca:2021wxt}.
Discontinuities in energy density that are too large drastically soften the equation of state (EOS) and quickly destabilize the twin branch, making it impossible to accommodate massive pulsars.
This bound can be evaded and would even allow for stronger phase transitions if the maximum mass resides on the normal nucleonic branch, and the third-generation stars are lighter than the most massive nucleonic stars (hence ``dwarf'').
However, these dwarfs are generally dismissed as unphysical as no known processes within the standard model appear to be capable of producing them.
In this series of papers we aim to examine such unconventional scenarios in light of possible interplay with dark sector physics, and this letter shall focus on their implications for neutron star observables. 
In particular, we will show that the recent multimessenger observations might have already hinted at their existence, a hypothesis that future data will either strengthen or disapprove
given the plethora of testable implications it entails.

\begin{figure}
	\includegraphics[width=0.98\linewidth]{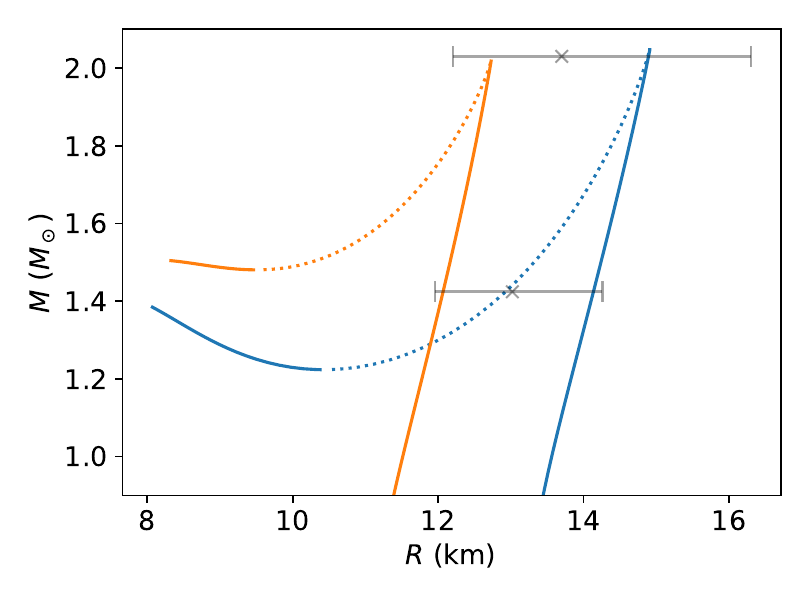}
	\caption{Mass-radius relations for two examples supporting dwarf twins.  Unstable branches are shown as dotted lines and are not physical. 
	The recent NICER measurements of the radii and $68\%$ confidence intervals for pulsars J0030+0451~\cite{Miller:2019cac} and J0740+6620~\cite{Miller:2021qha} 
	are shown in gray.
	Note that normal branches predicting larger radii do not necessarily lead to less massive dwarf branches, though their associated dwarf branches are more likely to support wider ranges of masses.}
	\label{fig:mr}
\end{figure}

The mass radius (MR) relations for two such examples are shown in \cref{fig:mr}.
While nucleonic branches support a wide range of masses from $\sim\msol$ to maximum values greater than $2~\msol$, dwarf branches only accommodate a fairly narrow interval typically less than $\sim0.4~\msol$. 
For concreteness, we shall refer to the maximum mass of the dwarf twins as the TOV limit, 
and that on the normal branch as the local maximum. 
As will be discussed below, although the TOV limit can be anywhere between $\sim 1-2.5\msol$, scenarios of astrophysical interest as suggested by GW170817 are those in the range $M_\tov\sim1.2-1.7~\msol$.
We have verified that the dwarf neutron stars are stable against radial perturbations~\cite{Bardeen_1966}, and that the empirical criterion first developed in \cite{1965gtgc.book.....H} applies to dwarf branches \footnote{That every counterclockwise bend of the MR curve (at extrema) adds an unstable mode and each clockwise turn removes one is proved in \cite{Had_i__2021} for polytropic EOSs.}.
This is unsurprising since dwarf twins are just a special type of twins.

Predictions for neutron star sizes are also rather unusual in scenarios accommodating dwarf twins.
On one hand, normal stars generally have large radii in the range $\sim12-16$ km, and are bigger in size for more massive stars.
On the other, dwarf stars are compact and have small radii typically in the range $\sim8-12$ km.
Some of the smallest radii in  fact breach the lower bound conjectured in ~\cite{Rhoades:1974fn} based on the maximally soft EOS.
Additionally, due to the large negative slope of MR relations for dwarf branches, 
the radii of dwarf stars may vary up to $\sim3$ km on each branch despite the tightly constrained masses.
These idiosyncratic features render the existence of dwarf twins an interesting hypothesis that can be tested by observations.\\

There are long-standing efforts at sizing neutron stars based on their thermal radiations.
Although earlier attempts are shown to be prone to large systematic uncertainties
~\cite{Catuneanu_2013,Heinke_2014}, recent proposals utilizing modulations in the surface brightness for calibration demonstrated encouraging potential in circumventing these issues~\cite{Miller_2015}.
Motivated by this promise the NICER mission was launched and has thus far produced two measurements. 
For the $1.44 \msol$ pulsar J0030+0451 its radius along with $68\%$ confidence intervals (CIs) are found to be $13.02^{+1.24}_{-1.06}$ km~\cite{Miller:2019cac}; for the $2.08\msol$ pulsar J0740+6620, $R=13.7^{+2.6}_{-1.5}$ km~\cite{Miller:2021qha}
\footnote{from analyses by the Maryland group. The Amsterdam group found similar results with tighter error bars~\cite{Riley:2019yda,Riley:2021pdl}.}. 
These pulsars are most likely normal stars on nucleonic branches.

Additional probes of neutron star radii have recently become possible thanks to the detection of gravitational waves from the binary neutron star merger event GW170817~\cite{Abbott:2017aa,Abbott:2018exr,Abbott:2018wiz}.
The modest value of tidal deformability, a parameter that is proportional to the fifth power of neutron star radius~\cite{Hinderer:2007mb},  extracted from waveforms ~\cite{Flanagan:2007ix} disfavors large radii for the stars involved.
For instance, the resulting constraint on the radius of $1.4~\msol$ stars is found to be $R_{1.4\msol}=11.0^{+0.9}_{-0.6}$ km at the $90\%$ level in \cite{Capano:2019eae}.
While it is still compatible with the $2\sigma$ NICER posteriors, tensions are building up.
We now show this discrepancy is resolved if 
either the primary (scenario I) or the secondary (scenario II) component in GW170817 is a dwarf twin, or if both components are dwarfs (scenario III). 

\begin{figure}
	\includegraphics[width=0.98\linewidth]{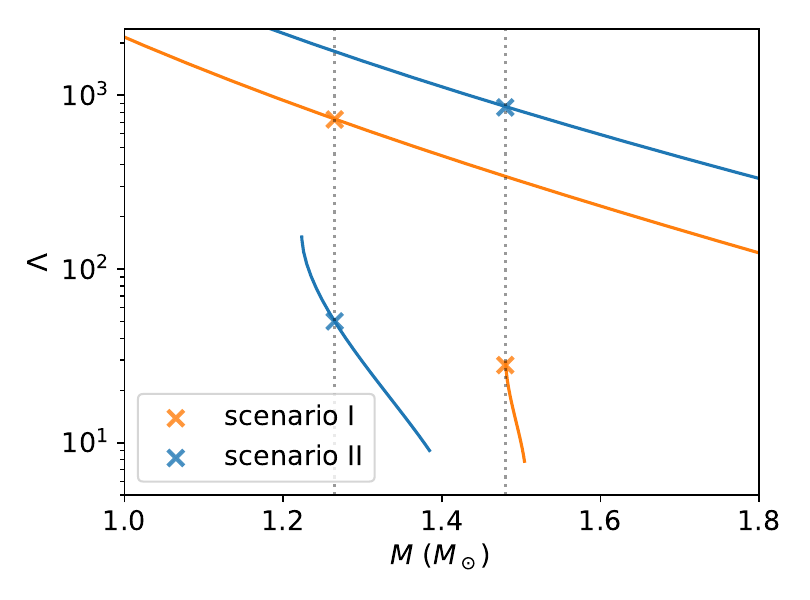}
	\caption{Tidal deformability $\Lambda$ of individual stars for MR relations shown in \cref{fig:mr}. For the central value $q=0.85$ inferred from GW170817 (which corresponds to $M_1=1.48\msol$ and $M_2=1.26\msol$), the blue (orange) curve is only compatible with scenario II (I), which predicts binary deformability $\widetilde\Lambda=550$ ($280$). These values are compatible with the bound $\widetilde\Lambda\lesssim600$ reported in the literature, despite predicting $R_{1.4\msol}\gtrsim12$ km on the nucleonic branch.
	Assuming only stars  on the normal branches are involved, the binary deformability for the blue and orange curves would be $\widetilde\Lambda=1200$ and $\widetilde\Lambda=490$ respectively at $q=0.85$.
	}\label{fig:lam}
\end{figure}

To demonstrate these possibilities we take central values of the inferred mass ratio $q\equiv M_2/M_1=0.85$ ($M_2\leq M_1$) and the chirp mass $\mchirp\equiv\frac{(M_1M_2)^{3/5}}{(M_1+M_2)^{1/5}}=1.188~\msol$ in \cite{Abbott:2017aa}, and form binaries for the MR relations shown in \cref{fig:mr}.
Their predictions for the tidal deformability $\Lambda$ are shown in \cref{fig:lam}.
The parameter that the waveform directly depends on is the binary tidal deformability $\widetilde\Lambda$ (see e.g. Supplemental Material \cite{suppl}), 
an averaged measure of $\Lambda$ for both the primary (1) and secondary (2) components. For $q=0.85$, it is given by
\begin{align}
\widetilde\Lambda(q=0.85) &\approx \Lambda_2\left(0.38 + 0.64\frac{\Lambda_1}{\Lambda_2}\right)\quad \text{scenario I: $\Lambda_1 < \Lambda_2$}\label{eq:case1}\\
&= \Lambda_1\left(0.64+ 0.38\frac{\Lambda_2}{\Lambda_1}\right)\quad \text{scenario II:  $\Lambda_2 < \Lambda_1$}\label{eq:case2}.
\end{align}

Since $\Lambda$ is (much) smaller for dwarfs than for normal stars, 
the binary deformability $\widetilde{\Lambda}$ is always dominated by stars on the normal branch.
\Cref{eq:case1,eq:case2} are arranged to reflect this.
It follows that for typical values of $\min\{\Lambda_1,\Lambda_2\}/\max\{\Lambda_1,\Lambda_2\}\lesssim0.1$, 
$\widetilde\Lambda\lesssim 0.45 \Lambda_2$ in scenario I, and $\widetilde\Lambda\lesssim 0.68 \Lambda_1$ in scenario II.
Taking the upper bound $\widetilde{\Lambda}\lesssim 600$ from GW170817 ~\cite{De:2018uhw,Capano:2019eae},
we find $\Lambda_2\equiv\Lambda(1.26\msol)\lesssim1300$ in scenario I, and $\Lambda_1\equiv\Lambda(1.46\msol)\lesssim890$ in scenario II. 
These bounds are about $40-80\%$ higher than would have been obtained assuming standard scenarios without dwarfs, where $\Lambda_1\leq\Lambda_2$, and
$\Lambda_2/\Lambda_1\simeq 0.35-0.45$ 
at $q=0.85$.
Using the correlation $\Lambda\propto (R/M)^{5.5}$ exhibited by stars on normal branches~\cite{suppl},
we find the relaxed upper bound on deformability raises the upper bound on radius by about a kilometer.
A generalized discussion for arbitrary mass ratios is presented in Supplemental Material \cite{suppl}.

We note that the upper bound $\widetilde\Lambda\lesssim500-600$ reported in the literature have either explicitly or implicitly 
assumed that the radii of neutron stars are mostly constant in the mass range $\sim1.2-1.8\msol$, a condition severely violated by the presence of dwarfs. 
One simple fix is to be completely agnostic and drop the common EOS assumption, as has been done in \cite{Abbott:2017aa}, where it is found that $\widetilde\Lambda\lesssim800$.
This suggests GW170817 could be compatible with higher values of $\widetilde\Lambda\sim 600-800$ if it were exotic.

Ideally, one would perform a Bayesian analysis of the gravitational wave data with a prior that incorporates dwarfs.
However, the narrow mass range supported by dwarf branches presents a technical challenge. 
Implementations based on existing frameworks would severely bias against dwarfs, as it is highly unlikely for samplers to settle on them. 
Improved sampling schemes are under development and will be reported later.
Here, we take a first step by analyzing an ensemble containing $\sim5000$ samples that all  support dwarf twins in the mass range of interest. For each sample we randomly generate $q$ in the range $[0.7,1.0]$ until scenario I (and scenario II in a separate analysis) can be constructed. For scenario I the resulting $\widetilde\Lambda\in[360, 2030]$ and for scenario II $\widetilde\Lambda\in[230, 2100]$ 
\footnote{The large $\widetilde\Lambda_\mathrm{min}$ in the prior reflects the large radii on normal branches $R\gtrsim12$ km, and could further shift the posterior in favor of larger $\widetilde{\Lambda}$ in Bayesian analyses accounting for dwarfs.}.
Upon imposing $\widetilde\Lambda\lesssim800$ (600) we find $R_{1.4\msol}\lesssim 14$ km (13.5 km) on the normal branch, consistent with findings from NICER.

Scenario III predicts even lower values of $\widetilde{\Lambda}$.
Because $\Lambda$ does not exceed $\sim 400$ on dwarf branches,  $\widetilde\Lambda$ of the binary would not surpass this value either.
However, GW170817 also yielded a lower bound $\widetilde{\Lambda}\gtrsim 60-80$ ~\cite{De:2018uhw,Capano:2019eae}.
This disfavors scenario III in which both $R_1$ and $R_2$ are small, or equivalently small values of $R_2$, since $R_1\leq R_2$ on all dwarf branches.
We repeat the analysis described above for scenario III, and find that 
$\widetilde{\Lambda}\geq 80$ roughly translates to $R_2\gtrsim 10$ km.
Taking into account $R_{1.4\msol}\lesssim14$ km from NICER,
this bound suggests that $M_2$ is very close to the minimum mass on dwarf branches.
Indeed, if future NICER observations reveal $R_{1.4\msol}\lesssim12$ km, 
a lower bound on $\widetilde\Lambda\gtrsim 100$ derived from GW170817-like events would not admit dwarfs as both components.

Binary neutron star mergers involving dwarf twins may also yield higher amounts of ejecta  which help explain the light curve of AT2017gfo~\cite{Nicholl:2017ahq, Kasen:2017sxr}.
Previous works found indications that smaller radii may produce more dynamical ejecta due to closer proximity at impact~\cite{Bauswein:2013yna,Dietrich:2016fpt}, although no clear correlations emerge~\cite{Radice:2018pdn}. 
Dwarf neutron stars are the most compact configurations and would likely help produce adequate dynamical ejecta responsible for the blue component of the kilonova light curve.
Using the empirical formula proposed in \cite{Dietrich:2016fpt} and the fitted parameters reported in \cite{Radice:2018pdn}, we find that for scenarios I and II the ejecta mass are about $20\%-80\%$ higher compared to canonical scenarios assuming normal neutron stars with radii $\sim10-11$ km, and that in scenario III the enhancement is about a factor of $\sim2-4$.
However, we note that the empirical fit is known to be crude and simulations are required to obtain reliable estimates of the dynamical ejecta in the presence of dwarfs.

Furthermore, in both scenarios I and II, the non-dwarf components are puffier than ordinary neutron stars and hence more susceptible to tidal deformations. This is another probable boost to mass shedding.
Finally,  due to the rapid stiffening inside the cores of dwarfs (to be discussed below), in all three scenarios stronger shockwaves are expected which further enhance the ejecta, and are likely the key to preventing the remnant from collapsing promptly to black holes (BHs).
A comprehensive view of mergers that is required to reveal nucleosynthesis yields and kilonova light curves is only possible through general relativistic hydrodynamic simulations and will be reported in future work.

Yet another sign in favor of dwarf twins comes from terrestrial laboratories.
Recently, the PREX-II experiment reported large symmetry energy $S$ and slope $L$ based on the measurement of neutron skin thickness of $^{208}\mathrm{Pb}$~\cite{PREX:2021umo}.
Assuming the parabolic expansion is valid and can be trusted for densities above $n_0$, this result favors large neutron star radii $R_{1.4\msol}\gtrsim13$ km~\cite{Reed:2021nqk}. 
This is in agreement with predictions for nucleonic stars on the normal branch.
However, we note that the correlation between the neutron skin measurements and astrophysical observables might not be strong as the former is most sensitive to the EOS at densities around $\sim0.7n_0$ whereas the relevant physics for dwarf twins occurs above $n_0$ (see below).
Dwarf twins cannot be ruled out even if lower values of $S$ and $L$ are reported in the future.\\

\begin{figure}
	\includegraphics[width=0.98\linewidth]{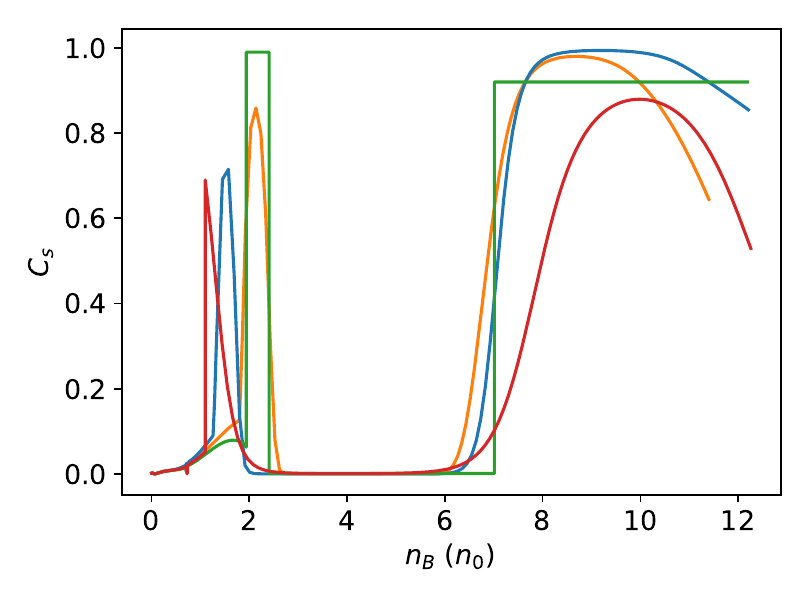}
	\caption{Speed of sound squared for a few EOSs underlying dwarf branches of astrophysical interest. The blue and orange EOSs are behind the results shown in the same colors in \cref{fig:mr,fig:lam}. The necessary condition for dwarfs is the presence of two peaks. Their shapes are relatively unimportant compared to their heights. The simple three-piece constant $C_s$ parameterization (green) is instrumental in deriving bounds on the peak prominence reported in the main text.}\label{fig:cs}
\end{figure}

The underlying equations of state that give rise to dwarfs share intriguing similarities.
The speed of sound squared $C_s$ of these EOSs increases rapidly between $n_0$ and $2n_0$ and drops sharply to values close to zero, maintaining a prolonged period of phase transition before shooting up again to reach a second peak.
A few examples are shown in \cref{fig:cs}.
The height of the first peak in $C_s$ directly controls the slope of MR relations (hence radii) on the normal branch.
Assuming chiral effective field theory ($\ceft$) is valid up to $n_0$ ($2n_0$)
, in order to accommodate both two-solar-mass stars and dwarf twin branches with $M_\tov\lesssim 1.5~M_\odot$ 
, this first peak is required to be at least $\sim0.4$ ($0.7$), leading to $R_{1.4\msol}\gtrsim 13.0$ km ($12.0$ km) and $R_{2.0\msol}\gtrsim 13.8$ km ($12.7$ km) on the normal branch.
Furthermore, jumps in energy density associated with the first-order phase transition are typically huge $\Delta\eden\sim500-1200$ MeV/fm$^3$ ($\Delta\eden/\eden_\mathrm{onset}\sim2-5$), and the second peak in $C_s$ needs to exceed $\sim0.6$ ($0.8$ if $\ceft$ is valid up to $2n_0$).

The strong phase transition pushes central densities at the TOV limit beyond $\eden_\mathrm{TOV}\simeq2.5-2.7$ GeV, exceeding the upper bound postulated in \cite{Drischler:2020fvz}.
The high densities inside dwarf suggest the underlying EOSs might be at odds with perturbative QCD (pQCD) predictions as these EOSs may not be able to reach the pQCD EOS in ways consistent with thermodynamics ~\cite{Zhou:2023zrm}. 
However, the phase transition also results in higher pressure at given chemical potentials, rendering pQCD constraints almost impotent.
For the ensemble of EOSs discussed above we found only around $10\%$ are at risk of being ruled out in the most constraining scenario.
The one exception is that if future observations disfavor large radii for normal stars, the maximum of $C_s$ needs to be higher to support dwarf branches, and this could make pQCD considerations relevant ~\cite{Zhou:2023zrm}.

There are well-motivated conjectures about QCD that may explain this unique family of EOSs.
The rapid increase in $C_s$ that leads to the first peak is likely an indication of the quarkyonic phase proposed in~\cite{McLerran:2018hbz}.
The second peak in $C_s$ above $\sim0.6$ suggests the phase of matter at these densities is likely dominated by non-perturbative phenomena. For instance, it might be a color superconductor ~\cite{Rajagopal:2000wf,Alford:2007xm} with a gap $\Delta\sim100$ MeV that is highly density-dependent.
However, due to the lack of non-perturbative descriptions we cannot provide a concrete description for the conjectures made here.
It is also possible that the peaks in $C_s$ are attributed to currently unknown physics.
If dwarf twins are confirmed by astrophysical observations, the peculiar features underlying their EOSs will deepen our understanding of cold and dense QCD.

As more and better observations are anticipated from LIGO and NICER, the existence of dwarf twins will be tested in the coming years.
On one hand, if the tension between refined and reliable measurements of neutron star radii from gravitational waves and from NICER persists,
or if these measurements exhibit large variability in radii or tidal deformabilites,
dwarf twin branches are probably the best explanation \footnote{While the dark halo scenario proposed in ~\cite{Nelson:2018xtr} could also cause large variability in tidal deformabilities, the radii of baryonic surfaces are not appreciably affected by dark halos. Furthermore, variations induced by dark halos are most sensitive to the total amount of dark matter which might be independent of stellar masses, whereas here the variations in tidal deformability  strongly correlate with masses.}.
On the other hand, if pulsars heavier than  $\sim2.5 ~\msol$ are found, dwarf twins with $M_\tov\lesssim 1.6 ~\msol$ will be effectively ruled out assuming $\ceft$ is valid up to $n_0$.

Because the masses of dwarf twins are less than the local maximum on normal branches, it is unlikely to produce dwarf twins via accretions onto normal neutron stars or collapses of massive stellar cores~\footnote{There is a tantalizing possibility that (certain special) core-collapse supernovae are not halted by the nuclear repulsive forces (the first peak in $C_s$), proceeding beyond the first-order transition and are reversed by the second peak in $C_s$ (which predicts  much higher bulk modulus $K=\eden C_s$ since energy density $\eden$ can be an order-of-magnitude higher than that in the nucleonic phase), thereby explaining dwarf twins within the Standard Model. This scenario likely requires considerable fine-tuning and might still rely on BSM physics for the rather extreme initial conditions. We defer to a subsequent work to scrutinize this possibility.}.
One possible formation scenario is through core-collapse of their nucleonic twins, a process we shall refer to as ``dwarfnova''.
We dedicate the subsequent paper to the mechanism and dynamics of such exotic events, and outline here their astrophysical implications and observable signatures that are indicative of the existence of dwarfs.

To begin with, the majority of the gravitational binding energy released during dwarfnovae are expected to be carried away by neutrinos.
Within a fully general-relativistic framework the energy budget is found to be $\sim0.5-5\times10^{53}$ ergs.
Although ordinary core-collapse supernovae are expected to release a similar amount of neutrinos with analogous spectra,
the key difference is the absence of neutronization bursts for dwarfnovae, as the initial configurations are already neutron-rich.
This will help the next-generation neutrino detectors such as DUNE to distinguish dwarfnovae from supernovae, as they can detect both $\bar{\nu}_e$ and $\nu_e$ and are anticipated to pick up thousands of neutrinos for  an implosion in our own galaxy~\cite{DUNE:2022aul}.
Additionally, assuming the neutron star mass function is not narrowly peaked, dwarfnovae will likely produce a considerable population of stellar mass BHs, since BHs are the only destiny for normal stars 
without dwarf counterparts. And if some of these BHs are hosted in binaries, 
even larger variability in $\widetilde{\Lambda}$ are expected in future gravitational wave detections .

During the conception and completion of this work the author is supported 
by the Institute for Nuclear Theory Grant No. DE-FG02-00ER41132 from the Department of Energy and Grant No. PHY-1430152 (JINA Center for the Evolution of the Elements, and
by NSF PFC 2020275 (Network for Neutrinos, Nuclear Astrophysics, and Symmetries (N3AS)).


%

\clearpage
\newpage
\section*{Supplemental Material: Binary tidal deformability}

The ``reduced'' tidal deformability for a binary consisting of components $1$ and $2$ is defined as
\begin{equation}
\widetilde\Lambda = \frac{16}{13}\frac{(12q+1)\Lambda_1+(12+q)q^4 \Lambda_2}{(1+q)^5},
\end{equation}
where $q=M_2/M_1\leq1$ is the mass ratio, and $\Lambda_{1,2}$ are the tidal deformability of each component stars.
For mass ratios close to $1$  ($q\gtrsim0.8$) an expansion in $(1-q)$ is useful and leads to
 \begin{align}
 \frac{\widetilde\Lambda}{\Lambda_1} &= \frac{1}{2}\lrp{1+\frac{\Lambda_2}{\Lambda_1}}+\frac{41}{52}\lrp{1-\frac{\Lambda_2}{\Lambda_1}}(1-q) + \mathcal{O}\lrb{(1-q)^2}\\
&\approx(1.288-0.788q)-(0.288-0.788q)\frac{\Lambda_2}{\Lambda_1}.
 \end{align}

\begin{figure}[!htbp]
	\includegraphics[width=0.98\linewidth]{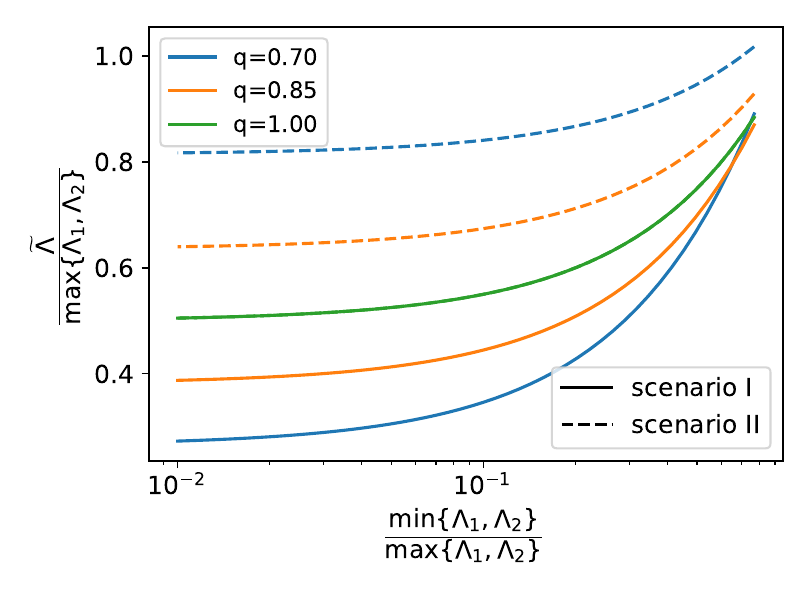}
	\caption{Suppression of $\widetilde\Lambda$ by dwarf twins in scenarios I and II.
	In scenarios I and II, $\min\{\Lambda_1,\Lambda_2\}/\max\{\Lambda_1,\Lambda_2\}\lesssim 0.1$, whereas 
	in the conventional cases without dwarfs $\min\{\Lambda_1,\Lambda_2\}/\max\{\Lambda_1,\Lambda_2\}\gtrsim 0.5$.
	We note that only scenario I has analogies in the standard cases as heavier stars always have lower values of deformability $\Lambda_2\geq\Lambda_1$  on each individual branch.
	}\label{fig:Lratio}
\end{figure}

\begin{figure}[!htbp]
	\includegraphics[width=0.98\linewidth]{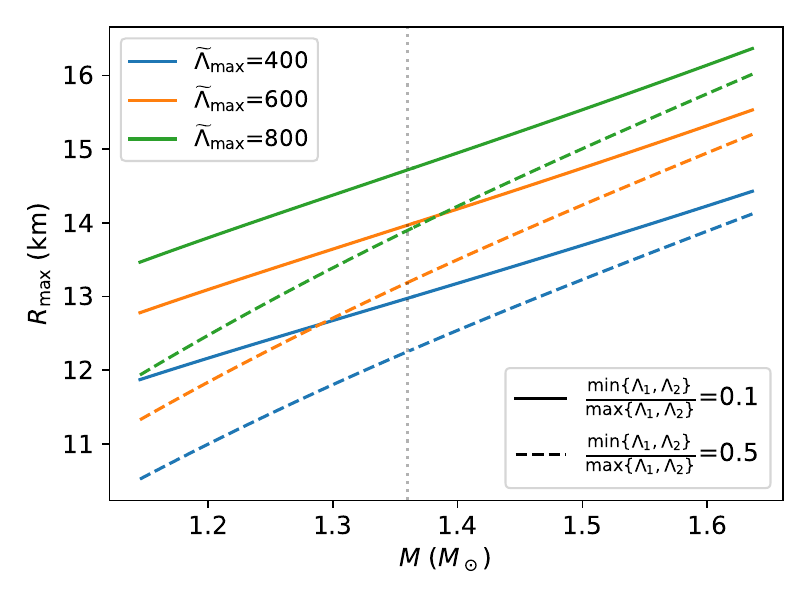}
	\caption{Translating upper bounds on $\widetilde{\Lambda}$ to those on radii using the power law fit \cref{eq:fit}. The dotted black line indicates the binary of mass ratio $q=1$. Curves to the left (right) of this line assume that the primary (secondary) component is a dwarf twin, and are obtained using \cref{eq:sc1} (\cref{eq:sc2}). 
	The solid curves assume $\frac{\min\{\Lambda_1,\Lambda_2\}}{\max\{\Lambda_1,\Lambda_2\}}=0.1$, a typical value (on the larger end) when dwarfs are involved.
	}\label{fig:rmax}
\end{figure}

Since $\widetilde\Lambda$ is dominated by the larger of the $\Lambda_{1,2}$, it is revealing to inspect its ratio to $\max\{\Lambda_1, \Lambda_2\}$.  In scenarios I and II, this ratio is given by
\begin{align}
\frac{\widetilde\Lambda}{\Lambda_2} &= \frac{16}{13}\frac{(12q+1)(\Lambda_1/\Lambda_2)+(12+q)q^4 }{(1+q)^5}\quad \text{scenario I,}\label{eq:sc1}\\
\frac{\widetilde\Lambda}{\Lambda_1}&= \frac{16}{13}\frac{(12q+1)+(12+q)q^4 (\Lambda_2/\Lambda_1)}{(1+q)^5}\quad \text{scenario II,}\label{eq:sc2}
\end{align}
respectively. \Cref{fig:Lratio} shows these relations for typical values of $\Lambda_i/\Lambda_j$ for binaries involving dwarf neutron stars. Even though scenario I appears to be favored as it is more potent in suppressing contributions from $\Lambda$ on the nucleonic branch, they accommodate smaller mass on the nucleonic branch and therefore predict higher $\max\{\Lambda_1,\Lambda_2\}$.
For the ~5k samples supporting dwarf twins examined in the main text, the resulting $\widetilde{\Lambda}$ appears to be similar for both cases.

We now examine the implications for neutron star radii.
For the 5k samples discussed above, we find that stars on the normal branch admit the following empirical relations (we adopt the natural unit in which $G=c=1$ throughout the manuscript)
\begin{equation}\label{eq:fit}
\Lambda\approx k \lrp{\frac{R}{M}}^{a},
\end{equation}
where $k\in [0.01,0.05]$ and $a\in[5.2, 5.8]$.
Below we take the best fit $k\approx0.25$ and $a\approx 5.5$.
This exponent is on the lower end among those in the so-called universality relations because the slopes of MR relations for normal branches here are positive and typically large. 
The resulting bound on radii is shown in \cref{fig:rmax}. 
We note that the dashed lines $\frac{\min\{\Lambda_1,\Lambda_2\}}{\max\{\Lambda_1,\Lambda_2\}}=0.5$ aims to mimic standard scenarios in the absence of dwarfs. They overestimate $R_\mathrm{max}$ because
the best fit exponent for ordinary stars whose radii may be as small as $\sim9$ km are higher than $a\sim5.5$. 
This will tilt the curves more and resulting in tighter bounds on $R_\mathrm{max}$.
Nevertheless, these crude estimates for arbitrary mass ratios again show that the  presence of dwarfs in GW170817 lowers the upper bound on radii by at least a kilometer, thereby reconciling the apparent discrepancy between NICER and GW170817.


\begin{thebibliography}{39}%
\makeatletter
\providecommand \@ifxundefined [1]{%
 \@ifx{#1\undefined}
}%
\providecommand \@ifnum [1]{%
 \ifnum #1\expandafter \@firstoftwo
 \else \expandafter \@secondoftwo
 \fi
}%
\providecommand \@ifx [1]{%
 \ifx #1\expandafter \@firstoftwo
 \else \expandafter \@secondoftwo
 \fi
}%
\providecommand \natexlab [1]{#1}%
\providecommand \enquote  [1]{``#1''}%
\providecommand \bibnamefont  [1]{#1}%
\providecommand \bibfnamefont [1]{#1}%
\providecommand \citenamefont [1]{#1}%
\providecommand \href@noop [0]{\@secondoftwo}%
\providecommand \href [0]{\begingroup \@sanitize@url \@href}%
\providecommand \@href[1]{\@@startlink{#1}\@@href}%
\providecommand \@@href[1]{\endgroup#1\@@endlink}%
\providecommand \@sanitize@url [0]{\catcode `\\12\catcode `\$12\catcode
  `\&12\catcode `\#12\catcode `\^12\catcode `\_12\catcode `\%12\relax}%
\providecommand \@@startlink[1]{}%
\providecommand \@@endlink[0]{}%
\providecommand \url  [0]{\begingroup\@sanitize@url \@url }%
\providecommand \@url [1]{\endgroup\@href {#1}{\urlprefix }}%
\providecommand \urlprefix  [0]{URL }%
\providecommand \Eprint [0]{\href }%
\providecommand \doibase [0]{https://doi.org/}%
\providecommand \selectlanguage [0]{\@gobble}%
\providecommand \bibinfo  [0]{\@secondoftwo}%
\providecommand \bibfield  [0]{\@secondoftwo}%
\providecommand \translation [1]{[#1]}%
\providecommand \BibitemOpen [0]{}%
\providecommand \bibitemStop [0]{}%
\providecommand \bibitemNoStop [0]{.\EOS\space}%
\providecommand \EOS [0]{\spacefactor3000\relax}%
\providecommand \BibitemShut  [1]{\csname bibitem#1\endcsname}%
\let\auto@bib@innerbib\@empty
\bibitem [{\citenamefont {Son}(1999)}]{Son:1998uk}%
  \BibitemOpen
  \bibfield  {author} {\bibinfo {author} {\bibfnamefont {D.~T.}\ \bibnamefont
  {Son}},\ }\bibfield  {title} {\bibinfo {title} {{Superconductivity by long
  range color magnetic interaction in high density quark matter}},\ }\href
  {https://doi.org/10.1103/PhysRevD.59.094019} {\bibfield  {journal} {\bibinfo
  {journal} {Phys. Rev. D}\ }\textbf {\bibinfo {volume} {59}},\ \bibinfo
  {pages} {094019} (\bibinfo {year} {1999})},\ \Eprint
  {https://arxiv.org/abs/hep-ph/9812287} {arXiv:hep-ph/9812287} \BibitemShut
  {NoStop}%
\bibitem [{\citenamefont {Rajagopal}\ and\ \citenamefont
  {Wilczek}(2000)}]{Rajagopal:2000wf}%
  \BibitemOpen
  \bibfield  {author} {\bibinfo {author} {\bibfnamefont {K.}~\bibnamefont
  {Rajagopal}}\ and\ \bibinfo {author} {\bibfnamefont {F.}~\bibnamefont
  {Wilczek}},\ }\bibinfo {title} {{The Condensed matter physics of QCD}},\ in\
  \href {https://doi.org/10.1142/9789812810458_0043} {\emph {\bibinfo
  {booktitle} {{At the frontier of particle physics. Handbook of QCD. Vol.
  1-3}}}},\ \bibinfo {editor} {edited by\ \bibinfo {editor} {\bibfnamefont
  {M.}~\bibnamefont {Shifman}}\ and\ \bibinfo {editor} {\bibfnamefont
  {B.}~\bibnamefont {Ioffe}}}\ (\bibinfo {year} {2000})\ pp.\ \bibinfo {pages}
  {2061--2151},\ \Eprint {https://arxiv.org/abs/hep-ph/0011333}
  {arXiv:hep-ph/0011333} \BibitemShut {NoStop}%
\bibitem [{\citenamefont {Alford}\ \emph {et~al.}(2008)\citenamefont {Alford},
  \citenamefont {Schmitt}, \citenamefont {Rajagopal},\ and\ \citenamefont
  {Sch\"afer}}]{Alford:2007xm}%
  \BibitemOpen
  \bibfield  {author} {\bibinfo {author} {\bibfnamefont {M.~G.}\ \bibnamefont
  {Alford}}, \bibinfo {author} {\bibfnamefont {A.}~\bibnamefont {Schmitt}},
  \bibinfo {author} {\bibfnamefont {K.}~\bibnamefont {Rajagopal}},\ and\
  \bibinfo {author} {\bibfnamefont {T.}~\bibnamefont {Sch\"afer}},\ }\bibfield
  {title} {\bibinfo {title} {{Color superconductivity in dense quark matter}},\
  }\href {https://doi.org/10.1103/RevModPhys.80.1455} {\bibfield  {journal}
  {\bibinfo  {journal} {Rev. Mod. Phys.}\ }\textbf {\bibinfo {volume} {80}},\
  \bibinfo {pages} {1455} (\bibinfo {year} {2008})},\ \Eprint
  {https://arxiv.org/abs/0709.4635} {arXiv:0709.4635 [hep-ph]} \BibitemShut
  {NoStop}%
\bibitem [{\citenamefont {Demorest}\ \emph {et~al.}(2010)\citenamefont
  {Demorest}, \citenamefont {Pennucci}, \citenamefont {Ransom}, \citenamefont
  {Roberts},\ and\ \citenamefont {Hessels}}]{Demorest:2010bx}%
  \BibitemOpen
  \bibfield  {author} {\bibinfo {author} {\bibfnamefont {P.}~\bibnamefont
  {Demorest}}, \bibinfo {author} {\bibfnamefont {T.}~\bibnamefont {Pennucci}},
  \bibinfo {author} {\bibfnamefont {S.}~\bibnamefont {Ransom}}, \bibinfo
  {author} {\bibfnamefont {M.}~\bibnamefont {Roberts}},\ and\ \bibinfo {author}
  {\bibfnamefont {J.}~\bibnamefont {Hessels}},\ }\bibfield  {title} {\bibinfo
  {title} {{Shapiro Delay Measurement of A Two Solar Mass Neutron Star}},\
  }\href {https://doi.org/10.1038/nature09466} {\bibfield  {journal} {\bibinfo
  {journal} {Nature}\ }\textbf {\bibinfo {volume} {467}},\ \bibinfo {pages}
  {1081} (\bibinfo {year} {2010})},\ \Eprint {https://arxiv.org/abs/1010.5788}
  {arXiv:1010.5788 [astro-ph.HE]} \BibitemShut {NoStop}%
\bibitem [{\citenamefont {Antoniadis}\ \emph {et~al.}(2013)\citenamefont
  {Antoniadis} \emph {et~al.}}]{Antoniadis:2013pzd}%
  \BibitemOpen
  \bibfield  {author} {\bibinfo {author} {\bibfnamefont {J.}~\bibnamefont
  {Antoniadis}} \emph {et~al.},\ }\bibfield  {title} {\bibinfo {title} {{A
  Massive Pulsar in a Compact Relativistic Binary}},\ }\href
  {https://doi.org/10.1126/science.1233232} {\bibfield  {journal} {\bibinfo
  {journal} {Science}\ }\textbf {\bibinfo {volume} {340}},\ \bibinfo {pages}
  {6131} (\bibinfo {year} {2013})},\ \Eprint {https://arxiv.org/abs/1304.6875}
  {arXiv:1304.6875 [astro-ph.HE]} \BibitemShut {NoStop}%
\bibitem [{\citenamefont {Romani}\ \emph {et~al.}(2021)\citenamefont {Romani},
  \citenamefont {Kandel}, \citenamefont {Filippenko}, \citenamefont {Brink},\
  and\ \citenamefont {Zheng}}]{Romani:2021xmb}%
  \BibitemOpen
  \bibfield  {author} {\bibinfo {author} {\bibfnamefont {R.~W.}\ \bibnamefont
  {Romani}}, \bibinfo {author} {\bibfnamefont {D.}~\bibnamefont {Kandel}},
  \bibinfo {author} {\bibfnamefont {A.~V.}\ \bibnamefont {Filippenko}},
  \bibinfo {author} {\bibfnamefont {T.~G.}\ \bibnamefont {Brink}},\ and\
  \bibinfo {author} {\bibfnamefont {W.}~\bibnamefont {Zheng}},\ }\bibfield
  {title} {\bibinfo {title} {{PSR J1810+1744: Companion Darkening and a Precise
  High Neutron Star Mass}},\ }\href {https://doi.org/10.3847/2041-8213/abe2b4}
  {\bibfield  {journal} {\bibinfo  {journal} {Astrophys. J. Lett.}\ }\textbf
  {\bibinfo {volume} {908}},\ \bibinfo {pages} {L46} (\bibinfo {year}
  {2021})},\ \Eprint {https://arxiv.org/abs/2101.09822} {arXiv:2101.09822
  [astro-ph.HE]} \BibitemShut {NoStop}%
\bibitem [{\citenamefont {Cromartie}\ \emph {et~al.}(2019)\citenamefont
  {Cromartie} \emph {et~al.}}]{NANOGrav:2019jur}%
  \BibitemOpen
  \bibfield  {author} {\bibinfo {author} {\bibfnamefont {H.~T.}\ \bibnamefont
  {Cromartie}} \emph {et~al.} (\bibinfo {collaboration} {NANOGrav}),\
  }\bibfield  {title} {\bibinfo {title} {{Relativistic Shapiro delay
  measurements of an extremely massive millisecond pulsar}},\ }\href
  {https://doi.org/10.1038/s41550-019-0880-2} {\bibfield  {journal} {\bibinfo
  {journal} {Nature Astron.}\ }\textbf {\bibinfo {volume} {4}},\ \bibinfo
  {pages} {72} (\bibinfo {year} {2019})},\ \Eprint
  {https://arxiv.org/abs/1904.06759} {arXiv:1904.06759 [astro-ph.HE]}
  \BibitemShut {NoStop}%
\bibitem [{\citenamefont {Fonseca}\ \emph {et~al.}(2021)\citenamefont {Fonseca}
  \emph {et~al.}}]{Fonseca:2021wxt}%
  \BibitemOpen
  \bibfield  {author} {\bibinfo {author} {\bibfnamefont {E.}~\bibnamefont
  {Fonseca}} \emph {et~al.},\ }\bibfield  {title} {\bibinfo {title} {{Refined
  Mass and Geometric Measurements of the High-mass PSR J0740+6620}},\ }\href
  {https://doi.org/10.3847/2041-8213/ac03b8} {\bibfield  {journal} {\bibinfo
  {journal} {Astrophys. J. Lett.}\ }\textbf {\bibinfo {volume} {915}},\
  \bibinfo {pages} {L12} (\bibinfo {year} {2021})},\ \Eprint
  {https://arxiv.org/abs/2104.00880} {arXiv:2104.00880 [astro-ph.HE]}
  \BibitemShut {NoStop}%
\bibitem [{\citenamefont {Miller}\ \emph {et~al.}(2019)\citenamefont {Miller}
  \emph {et~al.}}]{Miller:2019cac}%
  \BibitemOpen
  \bibfield  {author} {\bibinfo {author} {\bibfnamefont {M.~C.}\ \bibnamefont
  {Miller}} \emph {et~al.},\ }\bibfield  {title} {\bibinfo {title} {{PSR
  J0030+0451 Mass and Radius from $NICER$ Data and Implications for the
  Properties of Neutron Star Matter}},\ }\href
  {https://doi.org/10.3847/2041-8213/ab50c5} {\bibfield  {journal} {\bibinfo
  {journal} {Astrophys. J. Lett.}\ }\textbf {\bibinfo {volume} {887}},\
  \bibinfo {pages} {L24} (\bibinfo {year} {2019})},\ \Eprint
  {https://arxiv.org/abs/1912.05705} {arXiv:1912.05705 [astro-ph.HE]}
  \BibitemShut {NoStop}%
\bibitem [{\citenamefont {Miller}\ \emph {et~al.}(2021)\citenamefont {Miller}
  \emph {et~al.}}]{Miller:2021qha}%
  \BibitemOpen
  \bibfield  {author} {\bibinfo {author} {\bibfnamefont {M.~C.}\ \bibnamefont
  {Miller}} \emph {et~al.},\ }\bibfield  {title} {\bibinfo {title} {{The Radius
  of PSR J0740+6620 from NICER and XMM-Newton Data}},\ }\href@noop {} {\
  (\bibinfo {year} {2021})},\ \Eprint {https://arxiv.org/abs/2105.06979}
  {arXiv:2105.06979 [astro-ph.HE]} \BibitemShut {NoStop}%
\bibitem [{\citenamefont {Bardeen}\ \emph {et~al.}(1966)\citenamefont
  {Bardeen}, \citenamefont {Thorne},\ and\ \citenamefont
  {Meltzer}}]{Bardeen_1966}%
  \BibitemOpen
  \bibfield  {author} {\bibinfo {author} {\bibfnamefont {J.~M.}\ \bibnamefont
  {Bardeen}}, \bibinfo {author} {\bibfnamefont {K.~S.}\ \bibnamefont
  {Thorne}},\ and\ \bibinfo {author} {\bibfnamefont {D.~W.}\ \bibnamefont
  {Meltzer}},\ }\bibfield  {title} {\bibinfo {title} {A catalogue of methods
  for studying the normal modes of radial pulsation of general-relativistic
  stellar models},\ }\href {https://doi.org/10.1086/148791} {\bibfield
  {journal} {\bibinfo  {journal} {The Astrophysical Journal}\ }\textbf
  {\bibinfo {volume} {145}},\ \bibinfo {pages} {505} (\bibinfo {year}
  {1966})}\BibitemShut {NoStop}%
\bibitem [{\citenamefont {{Harrison}}\ \emph {et~al.}(1965)\citenamefont
  {{Harrison}}, \citenamefont {{Thorne}}, \citenamefont {{Wakano}},\ and\
  \citenamefont {{Wheeler}}}]{1965gtgc.book.....H}%
  \BibitemOpen
  \bibfield  {author} {\bibinfo {author} {\bibfnamefont {B.~K.}\ \bibnamefont
  {{Harrison}}}, \bibinfo {author} {\bibfnamefont {K.~S.}\ \bibnamefont
  {{Thorne}}}, \bibinfo {author} {\bibfnamefont {M.}~\bibnamefont {{Wakano}}},\
  and\ \bibinfo {author} {\bibfnamefont {J.~A.}\ \bibnamefont {{Wheeler}}},\
  }\href@noop {} {\emph {\bibinfo {title} {{Gravitation Theory and
  Gravitational Collapse}}}}\ (\bibinfo {year} {1965})\BibitemShut {NoStop}%
\bibitem [{\citenamefont {Had{\v{z}}i{\'{c}}}\ and\ \citenamefont
  {Lin}(2021)}]{Had_i__2021}%
  \BibitemOpen
  \bibfield  {author} {\bibinfo {author} {\bibfnamefont {M.}~\bibnamefont
  {Had{\v{z}}i{\'{c}}}}\ and\ \bibinfo {author} {\bibfnamefont
  {Z.}~\bibnamefont {Lin}},\ }\bibfield  {title} {\bibinfo {title} {Turning
  point principle for relativistic stars},\ }\href
  {https://doi.org/10.1007/s00220-021-04197-6} {\bibfield  {journal} {\bibinfo
  {journal} {Communications in Mathematical Physics}\ }\textbf {\bibinfo
  {volume} {387}},\ \bibinfo {pages} {729} (\bibinfo {year}
  {2021})}\BibitemShut {NoStop}%
\bibitem [{\citenamefont {Rhoades}\ and\ \citenamefont
  {Ruffini}(1974)}]{Rhoades:1974fn}%
  \BibitemOpen
  \bibfield  {author} {\bibinfo {author} {\bibfnamefont {C.~E.}\ \bibnamefont
  {Rhoades}, \bibfnamefont {Jr.}}\ and\ \bibinfo {author} {\bibfnamefont
  {R.}~\bibnamefont {Ruffini}},\ }\bibfield  {title} {\bibinfo {title}
  {{Maximum mass of a neutron star}},\ }\href
  {https://doi.org/10.1103/PhysRevLett.32.324} {\bibfield  {journal} {\bibinfo
  {journal} {Phys. Rev. Lett.}\ }\textbf {\bibinfo {volume} {32}},\ \bibinfo
  {pages} {324} (\bibinfo {year} {1974})}\BibitemShut {NoStop}%
\bibitem [{\citenamefont {Catuneanu}\ \emph {et~al.}(2013)\citenamefont
  {Catuneanu}, \citenamefont {Heinke}, \citenamefont {Sivakoff}, \citenamefont
  {Ho},\ and\ \citenamefont {Servillat}}]{Catuneanu_2013}%
  \BibitemOpen
  \bibfield  {author} {\bibinfo {author} {\bibfnamefont {A.}~\bibnamefont
  {Catuneanu}}, \bibinfo {author} {\bibfnamefont {C.~O.}\ \bibnamefont
  {Heinke}}, \bibinfo {author} {\bibfnamefont {G.~R.}\ \bibnamefont
  {Sivakoff}}, \bibinfo {author} {\bibfnamefont {W.~C.~G.}\ \bibnamefont
  {Ho}},\ and\ \bibinfo {author} {\bibfnamefont {M.}~\bibnamefont
  {Servillat}},\ }\bibfield  {title} {\bibinfo {title} {{MASS}/{RADIUS}
  {CONSTRAINTS} {ON} {THE} {QUIESCENT} {NEUTRON} {STAR} {IN} m13 {USING}
  {HYDROGEN} {AND} {HELIUM} {ATMOSPHERES}},\ }\href
  {https://doi.org/10.1088/0004-637x/764/2/145} {\bibfield  {journal} {\bibinfo
   {journal} {The Astrophysical Journal}\ }\textbf {\bibinfo {volume} {764}},\
  \bibinfo {pages} {145} (\bibinfo {year} {2013})}\BibitemShut {NoStop}%
\bibitem [{\citenamefont {Heinke}\ \emph {et~al.}(2014)\citenamefont {Heinke},
  \citenamefont {Cohn}, \citenamefont {Lugger}, \citenamefont {Webb},
  \citenamefont {Ho}, \citenamefont {Anderson}, \citenamefont {Campana},
  \citenamefont {Bogdanov}, \citenamefont {Haggard}, \citenamefont {Cool},\
  and\ \citenamefont {Grindlay}}]{Heinke_2014}%
  \BibitemOpen
  \bibfield  {author} {\bibinfo {author} {\bibfnamefont {C.~O.}\ \bibnamefont
  {Heinke}}, \bibinfo {author} {\bibfnamefont {H.~N.}\ \bibnamefont {Cohn}},
  \bibinfo {author} {\bibfnamefont {P.~M.}\ \bibnamefont {Lugger}}, \bibinfo
  {author} {\bibfnamefont {N.~A.}\ \bibnamefont {Webb}}, \bibinfo {author}
  {\bibfnamefont {W.~C.~G.}\ \bibnamefont {Ho}}, \bibinfo {author}
  {\bibfnamefont {J.}~\bibnamefont {Anderson}}, \bibinfo {author}
  {\bibfnamefont {S.}~\bibnamefont {Campana}}, \bibinfo {author} {\bibfnamefont
  {S.}~\bibnamefont {Bogdanov}}, \bibinfo {author} {\bibfnamefont
  {D.}~\bibnamefont {Haggard}}, \bibinfo {author} {\bibfnamefont {A.~M.}\
  \bibnamefont {Cool}},\ and\ \bibinfo {author} {\bibfnamefont {J.~E.}\
  \bibnamefont {Grindlay}},\ }\bibfield  {title} {\bibinfo {title} {Improved
  mass and radius constraints for quiescent neutron stars in $\omega$ cen and
  {NGC} 6397},\ }\href {https://doi.org/10.1093/mnras/stu1449} {\bibfield
  {journal} {\bibinfo  {journal} {Monthly Notices of the Royal Astronomical
  Society}\ }\textbf {\bibinfo {volume} {444}},\ \bibinfo {pages} {443}
  (\bibinfo {year} {2014})}\BibitemShut {NoStop}%
\bibitem [{\citenamefont {Miller}\ and\ \citenamefont
  {Lamb}(2015)}]{Miller_2015}%
  \BibitemOpen
  \bibfield  {author} {\bibinfo {author} {\bibfnamefont {M.~C.}\ \bibnamefont
  {Miller}}\ and\ \bibinfo {author} {\bibfnamefont {F.~K.}\ \bibnamefont
  {Lamb}},\ }\bibfield  {title} {\bibinfo {title} {{DETERMINING} {NEUTRON}
  {STAR} {PROPERTIES} {BY} {FITTING} {OBLATE}-{STAR} {WAVEFORM} {MODELS} {TO}
  x-{RAY} {BURST} {OSCILLATIONS}},\ }\href
  {https://doi.org/10.1088/0004-637x/808/1/31} {\bibfield  {journal} {\bibinfo
  {journal} {The Astrophysical Journal}\ }\textbf {\bibinfo {volume} {808}},\
  \bibinfo {pages} {31} (\bibinfo {year} {2015})}\BibitemShut {NoStop}%
\bibitem [{\citenamefont {Riley}\ \emph {et~al.}(2019)\citenamefont {Riley}
  \emph {et~al.}}]{Riley:2019yda}%
  \BibitemOpen
  \bibfield  {author} {\bibinfo {author} {\bibfnamefont {T.~E.}\ \bibnamefont
  {Riley}} \emph {et~al.},\ }\bibfield  {title} {\bibinfo {title} {{A $NICER$
  View of PSR J0030+0451: Millisecond Pulsar Parameter Estimation}},\ }\href
  {https://doi.org/10.3847/2041-8213/ab481c} {\bibfield  {journal} {\bibinfo
  {journal} {Astrophys. J.}\ }\textbf {\bibinfo {volume} {887}},\ \bibinfo
  {pages} {L21} (\bibinfo {year} {2019})},\ \Eprint
  {https://arxiv.org/abs/1912.05702} {arXiv:1912.05702 [astro-ph.HE]}
  \BibitemShut {NoStop}%
\bibitem [{\citenamefont {Riley}\ \emph {et~al.}(2021)\citenamefont {Riley}
  \emph {et~al.}}]{Riley:2021pdl}%
  \BibitemOpen
  \bibfield  {author} {\bibinfo {author} {\bibfnamefont {T.~E.}\ \bibnamefont
  {Riley}} \emph {et~al.},\ }\bibfield  {title} {\bibinfo {title} {{A NICER
  View of the Massive Pulsar PSR J0740+6620 Informed by Radio Timing and
  XMM-Newton Spectroscopy}},\ }\href {https://doi.org/10.3847/2041-8213/ac0a81}
  {\bibfield  {journal} {\bibinfo  {journal} {Astrophys. J. Lett.}\ }\textbf
  {\bibinfo {volume} {918}},\ \bibinfo {pages} {L27} (\bibinfo {year}
  {2021})},\ \Eprint {https://arxiv.org/abs/2105.06980} {arXiv:2105.06980
  [astro-ph.HE]} \BibitemShut {NoStop}%
\bibitem [{\citenamefont {et~al. LIGO Scientific~Collaboration}\ and\
  \citenamefont {Collaboration}(2017)}]{Abbott:2017aa}%
  \BibitemOpen
  \bibfield  {author} {\bibinfo {author} {\bibfnamefont {B.~P.~A.}\
  \bibnamefont {et~al. LIGO Scientific~Collaboration}}\ and\ \bibinfo {author}
  {\bibfnamefont {V.}~\bibnamefont {Collaboration}},\ }\bibfield  {title}
  {\bibinfo {title} {Gw170817: Observation of gravitational waves from a binary
  neutron star inspiral},\ }\bibfield  {journal} {\bibinfo  {journal} {Physical
  Review Letters}\ }\textbf {\bibinfo {volume} {119}},\ \href
  {https://doi.org/10.1103/PhysRevLett.119.161101}
  {10.1103/PhysRevLett.119.161101} (\bibinfo {year} {2017})\BibitemShut
  {NoStop}%
\bibitem [{\citenamefont {Abbott}\ \emph {et~al.}(2018)\citenamefont {Abbott}
  \emph {et~al.}}]{Abbott:2018exr}%
  \BibitemOpen
  \bibfield  {author} {\bibinfo {author} {\bibfnamefont {B.~P.}\ \bibnamefont
  {Abbott}} \emph {et~al.} (\bibinfo {collaboration} {LIGO Scientific,
  Virgo}),\ }\bibfield  {title} {\bibinfo {title} {{GW170817: Measurements of
  neutron star radii and equation of state}},\ }\href
  {https://doi.org/10.1103/PhysRevLett.121.161101} {\bibfield  {journal}
  {\bibinfo  {journal} {Phys. Rev. Lett.}\ }\textbf {\bibinfo {volume} {121}},\
  \bibinfo {pages} {161101} (\bibinfo {year} {2018})},\ \Eprint
  {https://arxiv.org/abs/1805.11581} {arXiv:1805.11581 [gr-qc]} \BibitemShut
  {NoStop}%
\bibitem [{\citenamefont {Abbott}\ \emph {et~al.}(2019)\citenamefont {Abbott}
  \emph {et~al.}}]{Abbott:2018wiz}%
  \BibitemOpen
  \bibfield  {author} {\bibinfo {author} {\bibfnamefont {B.~P.}\ \bibnamefont
  {Abbott}} \emph {et~al.} (\bibinfo {collaboration} {LIGO Scientific,
  Virgo}),\ }\bibfield  {title} {\bibinfo {title} {{Properties of the binary
  neutron star merger GW170817}},\ }\href
  {https://doi.org/10.1103/PhysRevX.9.011001} {\bibfield  {journal} {\bibinfo
  {journal} {Phys. Rev.}\ }\textbf {\bibinfo {volume} {X9}},\ \bibinfo {pages}
  {011001} (\bibinfo {year} {2019})},\ \Eprint
  {https://arxiv.org/abs/1805.11579} {arXiv:1805.11579 [gr-qc]} \BibitemShut
  {NoStop}%
\bibitem [{\citenamefont {Hinderer}(2008)}]{Hinderer:2007mb}%
  \BibitemOpen
  \bibfield  {author} {\bibinfo {author} {\bibfnamefont {T.}~\bibnamefont
  {Hinderer}},\ }\bibfield  {title} {\bibinfo {title} {{Tidal Love numbers of
  neutron stars}},\ }\href {https://doi.org/10.1086/533487} {\bibfield
  {journal} {\bibinfo  {journal} {Astrophys. J.}\ }\textbf {\bibinfo {volume}
  {677}},\ \bibinfo {pages} {1216} (\bibinfo {year} {2008})},\ \Eprint
  {https://arxiv.org/abs/0711.2420} {arXiv:0711.2420 [astro-ph]} \BibitemShut
  {NoStop}%
\bibitem [{\citenamefont {Flanagan}\ and\ \citenamefont
  {Hinderer}(2008)}]{Flanagan:2007ix}%
  \BibitemOpen
  \bibfield  {author} {\bibinfo {author} {\bibfnamefont {E.~E.}\ \bibnamefont
  {Flanagan}}\ and\ \bibinfo {author} {\bibfnamefont {T.}~\bibnamefont
  {Hinderer}},\ }\bibfield  {title} {\bibinfo {title} {{Constraining neutron
  star tidal Love numbers with gravitational wave detectors}},\ }\href
  {https://doi.org/10.1103/PhysRevD.77.021502} {\bibfield  {journal} {\bibinfo
  {journal} {Phys. Rev.}\ }\textbf {\bibinfo {volume} {D77}},\ \bibinfo {pages}
  {021502} (\bibinfo {year} {2008})},\ \Eprint
  {https://arxiv.org/abs/0709.1915} {arXiv:0709.1915 [astro-ph]} \BibitemShut
  {NoStop}%
\bibitem [{\citenamefont {Capano}\ \emph {et~al.}(2020)\citenamefont {Capano},
  \citenamefont {Tews}, \citenamefont {Brown}, \citenamefont {Margalit},
  \citenamefont {De}, \citenamefont {Kumar}, \citenamefont {Brown},
  \citenamefont {Krishnan},\ and\ \citenamefont {Reddy}}]{Capano:2019eae}%
  \BibitemOpen
  \bibfield  {author} {\bibinfo {author} {\bibfnamefont {C.~D.}\ \bibnamefont
  {Capano}}, \bibinfo {author} {\bibfnamefont {I.}~\bibnamefont {Tews}},
  \bibinfo {author} {\bibfnamefont {S.~M.}\ \bibnamefont {Brown}}, \bibinfo
  {author} {\bibfnamefont {B.}~\bibnamefont {Margalit}}, \bibinfo {author}
  {\bibfnamefont {S.}~\bibnamefont {De}}, \bibinfo {author} {\bibfnamefont
  {S.}~\bibnamefont {Kumar}}, \bibinfo {author} {\bibfnamefont {D.~A.}\
  \bibnamefont {Brown}}, \bibinfo {author} {\bibfnamefont {B.}~\bibnamefont
  {Krishnan}},\ and\ \bibinfo {author} {\bibfnamefont {S.}~\bibnamefont
  {Reddy}},\ }\bibfield  {title} {\bibinfo {title} {{Stringent constraints on
  neutron-star radii from multimessenger observations and nuclear theory}},\
  }\href {https://doi.org/10.1038/s41550-020-1014-6} {\bibfield  {journal}
  {\bibinfo  {journal} {Nature Astron.}\ }\textbf {\bibinfo {volume} {4}},\
  \bibinfo {pages} {625} (\bibinfo {year} {2020})},\ \Eprint
  {https://arxiv.org/abs/1908.10352} {arXiv:1908.10352 [astro-ph.HE]}
  \BibitemShut {NoStop}%
\bibitem [{sup(2023)}]{suppl}%
  \BibitemOpen
  \bibfield  {title} {\bibinfo {title} {{Supplemental Material}},\ }\href@noop
  {} {\  (\bibinfo {year} {2023})}\BibitemShut {NoStop}%
\bibitem [{\citenamefont {De}\ \emph {et~al.}(2018)\citenamefont {De},
  \citenamefont {Finstad}, \citenamefont {Lattimer}, \citenamefont {Brown},
  \citenamefont {Berger},\ and\ \citenamefont {Biwer}}]{De:2018uhw}%
  \BibitemOpen
  \bibfield  {author} {\bibinfo {author} {\bibfnamefont {S.}~\bibnamefont
  {De}}, \bibinfo {author} {\bibfnamefont {D.}~\bibnamefont {Finstad}},
  \bibinfo {author} {\bibfnamefont {J.~M.}\ \bibnamefont {Lattimer}}, \bibinfo
  {author} {\bibfnamefont {D.~A.}\ \bibnamefont {Brown}}, \bibinfo {author}
  {\bibfnamefont {E.}~\bibnamefont {Berger}},\ and\ \bibinfo {author}
  {\bibfnamefont {C.~M.}\ \bibnamefont {Biwer}},\ }\bibfield  {title} {\bibinfo
  {title} {Constraining the nuclear equation of state with {GW170817}},\ }\href
  {https://doi.org/10.1103/PhysRevLett.121.091102} {\bibfield  {journal}
  {\bibinfo  {journal} {Phys. Rev. Lett.}\ }\textbf {\bibinfo {volume} {121}},\
  \bibinfo {pages} {091102} (\bibinfo {year} {2018})},\ \Eprint
  {https://arxiv.org/abs/1804.08583} {arXiv:1804.08583 [astro-ph.HE]}
  \BibitemShut {NoStop}%
\bibitem [{\citenamefont {Nicholl}\ \emph {et~al.}(2017)\citenamefont {Nicholl}
  \emph {et~al.}}]{Nicholl:2017ahq}%
  \BibitemOpen
  \bibfield  {author} {\bibinfo {author} {\bibfnamefont {M.}~\bibnamefont
  {Nicholl}} \emph {et~al.},\ }\bibfield  {title} {\bibinfo {title} {{The
  Electromagnetic Counterpart of the Binary Neutron Star Merger LIGO/VIRGO
  GW170817. III. Optical and UV Spectra of a Blue Kilonova From Fast Polar
  Ejecta}},\ }\href {https://doi.org/10.3847/2041-8213/aa9029} {\bibfield
  {journal} {\bibinfo  {journal} {Astrophys. J. Lett.}\ }\textbf {\bibinfo
  {volume} {848}},\ \bibinfo {pages} {L18} (\bibinfo {year} {2017})},\ \Eprint
  {https://arxiv.org/abs/1710.05456} {arXiv:1710.05456 [astro-ph.HE]}
  \BibitemShut {NoStop}%
\bibitem [{\citenamefont {Kasen}\ \emph {et~al.}(2017)\citenamefont {Kasen},
  \citenamefont {Metzger}, \citenamefont {Barnes}, \citenamefont {Quataert},\
  and\ \citenamefont {Ramirez-Ruiz}}]{Kasen:2017sxr}%
  \BibitemOpen
  \bibfield  {author} {\bibinfo {author} {\bibfnamefont {D.}~\bibnamefont
  {Kasen}}, \bibinfo {author} {\bibfnamefont {B.}~\bibnamefont {Metzger}},
  \bibinfo {author} {\bibfnamefont {J.}~\bibnamefont {Barnes}}, \bibinfo
  {author} {\bibfnamefont {E.}~\bibnamefont {Quataert}},\ and\ \bibinfo
  {author} {\bibfnamefont {E.}~\bibnamefont {Ramirez-Ruiz}},\ }\bibfield
  {title} {\bibinfo {title} {{Origin of the heavy elements in binary
  neutron-star mergers from a gravitational wave event}},\ }\href
  {https://doi.org/10.1038/nature24453} {\bibfield  {journal} {\bibinfo
  {journal} {Nature}\ }\textbf {\bibinfo {volume} {551}},\ \bibinfo {pages}
  {80} (\bibinfo {year} {2017})},\ \Eprint {https://arxiv.org/abs/1710.05463}
  {arXiv:1710.05463 [astro-ph.HE]} \BibitemShut {NoStop}%
\bibitem [{\citenamefont {Bauswein}\ \emph {et~al.}(2013)\citenamefont
  {Bauswein}, \citenamefont {Goriely},\ and\ \citenamefont
  {Janka}}]{Bauswein:2013yna}%
  \BibitemOpen
  \bibfield  {author} {\bibinfo {author} {\bibfnamefont {A.}~\bibnamefont
  {Bauswein}}, \bibinfo {author} {\bibfnamefont {S.}~\bibnamefont {Goriely}},\
  and\ \bibinfo {author} {\bibfnamefont {H.~T.}\ \bibnamefont {Janka}},\
  }\bibfield  {title} {\bibinfo {title} {{Systematics of dynamical mass
  ejection, nucleosynthesis, and radioactively powered electromagnetic signals
  from neutron-star mergers}},\ }\href
  {https://doi.org/10.1088/0004-637X/773/1/78} {\bibfield  {journal} {\bibinfo
  {journal} {Astrophys. J.}\ }\textbf {\bibinfo {volume} {773}},\ \bibinfo
  {pages} {78} (\bibinfo {year} {2013})},\ \Eprint
  {https://arxiv.org/abs/1302.6530} {arXiv:1302.6530 [astro-ph.SR]}
  \BibitemShut {NoStop}%
\bibitem [{\citenamefont {Dietrich}\ and\ \citenamefont
  {Ujevic}(2017)}]{Dietrich:2016fpt}%
  \BibitemOpen
  \bibfield  {author} {\bibinfo {author} {\bibfnamefont {T.}~\bibnamefont
  {Dietrich}}\ and\ \bibinfo {author} {\bibfnamefont {M.}~\bibnamefont
  {Ujevic}},\ }\bibfield  {title} {\bibinfo {title} {{Modeling dynamical ejecta
  from binary neutron star mergers and implications for electromagnetic
  counterparts}},\ }\href {https://doi.org/10.1088/1361-6382/aa6bb0} {\bibfield
   {journal} {\bibinfo  {journal} {Class. Quant. Grav.}\ }\textbf {\bibinfo
  {volume} {34}},\ \bibinfo {pages} {105014} (\bibinfo {year} {2017})},\
  \Eprint {https://arxiv.org/abs/1612.03665} {arXiv:1612.03665 [gr-qc]}
  \BibitemShut {NoStop}%
\bibitem [{\citenamefont {Radice}\ \emph {et~al.}(2018)\citenamefont {Radice},
  \citenamefont {Perego}, \citenamefont {Hotokezaka}, \citenamefont {Fromm},
  \citenamefont {Bernuzzi},\ and\ \citenamefont {Roberts}}]{Radice:2018pdn}%
  \BibitemOpen
  \bibfield  {author} {\bibinfo {author} {\bibfnamefont {D.}~\bibnamefont
  {Radice}}, \bibinfo {author} {\bibfnamefont {A.}~\bibnamefont {Perego}},
  \bibinfo {author} {\bibfnamefont {K.}~\bibnamefont {Hotokezaka}}, \bibinfo
  {author} {\bibfnamefont {S.~A.}\ \bibnamefont {Fromm}}, \bibinfo {author}
  {\bibfnamefont {S.}~\bibnamefont {Bernuzzi}},\ and\ \bibinfo {author}
  {\bibfnamefont {L.~F.}\ \bibnamefont {Roberts}},\ }\bibfield  {title}
  {\bibinfo {title} {{Binary Neutron Star Mergers: Mass Ejection,
  Electromagnetic Counterparts and Nucleosynthesis}},\ }\href
  {https://doi.org/10.3847/1538-4357/aaf054} {\bibfield  {journal} {\bibinfo
  {journal} {Astrophys. J.}\ }\textbf {\bibinfo {volume} {869}},\ \bibinfo
  {pages} {130} (\bibinfo {year} {2018})},\ \Eprint
  {https://arxiv.org/abs/1809.11161} {arXiv:1809.11161 [astro-ph.HE]}
  \BibitemShut {NoStop}%
\bibitem [{\citenamefont {Adhikari}\ \emph {et~al.}(2021)\citenamefont
  {Adhikari} \emph {et~al.}}]{PREX:2021umo}%
  \BibitemOpen
  \bibfield  {author} {\bibinfo {author} {\bibfnamefont {D.}~\bibnamefont
  {Adhikari}} \emph {et~al.} (\bibinfo {collaboration} {PREX}),\ }\bibfield
  {title} {\bibinfo {title} {{Accurate Determination of the Neutron Skin
  Thickness of $^{208}$Pb through Parity-Violation in Electron Scattering}},\
  }\href {https://doi.org/10.1103/PhysRevLett.126.172502} {\bibfield  {journal}
  {\bibinfo  {journal} {Phys. Rev. Lett.}\ }\textbf {\bibinfo {volume} {126}},\
  \bibinfo {pages} {172502} (\bibinfo {year} {2021})},\ \Eprint
  {https://arxiv.org/abs/2102.10767} {arXiv:2102.10767 [nucl-ex]} \BibitemShut
  {NoStop}%
\bibitem [{\citenamefont {Reed}\ \emph {et~al.}(2021)\citenamefont {Reed},
  \citenamefont {Fattoyev}, \citenamefont {Horowitz},\ and\ \citenamefont
  {Piekarewicz}}]{Reed:2021nqk}%
  \BibitemOpen
  \bibfield  {author} {\bibinfo {author} {\bibfnamefont {B.~T.}\ \bibnamefont
  {Reed}}, \bibinfo {author} {\bibfnamefont {F.~J.}\ \bibnamefont {Fattoyev}},
  \bibinfo {author} {\bibfnamefont {C.~J.}\ \bibnamefont {Horowitz}},\ and\
  \bibinfo {author} {\bibfnamefont {J.}~\bibnamefont {Piekarewicz}},\
  }\bibfield  {title} {\bibinfo {title} {{Implications of PREX-2 on the
  Equation of State of Neutron-Rich Matter}},\ }\href
  {https://doi.org/10.1103/PhysRevLett.126.172503} {\bibfield  {journal}
  {\bibinfo  {journal} {Phys. Rev. Lett.}\ }\textbf {\bibinfo {volume} {126}},\
  \bibinfo {pages} {172503} (\bibinfo {year} {2021})},\ \Eprint
  {https://arxiv.org/abs/2101.03193} {arXiv:2101.03193 [nucl-th]} \BibitemShut
  {NoStop}%
\bibitem [{\citenamefont {Drischler}\ \emph {et~al.}(2021)\citenamefont
  {Drischler}, \citenamefont {Han}, \citenamefont {Lattimer}, \citenamefont
  {Prakash}, \citenamefont {Reddy},\ and\ \citenamefont
  {Zhao}}]{Drischler:2020fvz}%
  \BibitemOpen
  \bibfield  {author} {\bibinfo {author} {\bibfnamefont {C.}~\bibnamefont
  {Drischler}}, \bibinfo {author} {\bibfnamefont {S.}~\bibnamefont {Han}},
  \bibinfo {author} {\bibfnamefont {J.~M.}\ \bibnamefont {Lattimer}}, \bibinfo
  {author} {\bibfnamefont {M.}~\bibnamefont {Prakash}}, \bibinfo {author}
  {\bibfnamefont {S.}~\bibnamefont {Reddy}},\ and\ \bibinfo {author}
  {\bibfnamefont {T.}~\bibnamefont {Zhao}},\ }\bibfield  {title} {\bibinfo
  {title} {{Limiting masses and radii of neutron stars and their
  implications}},\ }\href {https://doi.org/10.1103/PhysRevC.103.045808}
  {\bibfield  {journal} {\bibinfo  {journal} {Phys. Rev. C}\ }\textbf {\bibinfo
  {volume} {103}},\ \bibinfo {pages} {045808} (\bibinfo {year} {2021})},\
  \Eprint {https://arxiv.org/abs/2009.06441} {arXiv:2009.06441 [nucl-th]}
  \BibitemShut {NoStop}%
\bibitem [{\citenamefont {Zhou}(2023)}]{Zhou:2023zrm}%
  \BibitemOpen
  \bibfield  {author} {\bibinfo {author} {\bibfnamefont {D.}~\bibnamefont
  {Zhou}},\ }\bibfield  {title} {\bibinfo {title} {{What does perturbative QCD
  really have to say about neutron stars}},\ }\href@noop {} {\  (\bibinfo
  {year} {2023})},\ \Eprint {https://arxiv.org/abs/2307.11125}
  {arXiv:2307.11125 [astro-ph.HE]} \BibitemShut {NoStop}%
\bibitem [{\citenamefont {McLerran}\ and\ \citenamefont
  {Reddy}(2019)}]{McLerran:2018hbz}%
  \BibitemOpen
  \bibfield  {author} {\bibinfo {author} {\bibfnamefont {L.}~\bibnamefont
  {McLerran}}\ and\ \bibinfo {author} {\bibfnamefont {S.}~\bibnamefont
  {Reddy}},\ }\bibfield  {title} {\bibinfo {title} {{Quarkyonic Matter and
  Neutron Stars}},\ }\href {https://doi.org/10.1103/PhysRevLett.122.122701}
  {\bibfield  {journal} {\bibinfo  {journal} {Phys. Rev. Lett.}\ }\textbf
  {\bibinfo {volume} {122}},\ \bibinfo {pages} {122701} (\bibinfo {year}
  {2019})},\ \Eprint {https://arxiv.org/abs/1811.12503} {arXiv:1811.12503
  [nucl-th]} \BibitemShut {NoStop}%
\bibitem [{\citenamefont {Nelson}\ \emph {et~al.}(2019)\citenamefont {Nelson},
  \citenamefont {Reddy},\ and\ \citenamefont {Zhou}}]{Nelson:2018xtr}%
  \BibitemOpen
  \bibfield  {author} {\bibinfo {author} {\bibfnamefont {A.}~\bibnamefont
  {Nelson}}, \bibinfo {author} {\bibfnamefont {S.}~\bibnamefont {Reddy}},\ and\
  \bibinfo {author} {\bibfnamefont {D.}~\bibnamefont {Zhou}},\ }\bibfield
  {title} {\bibinfo {title} {{Dark halos around neutron stars and gravitational
  waves}},\ }\href {https://doi.org/10.1088/1475-7516/2019/07/012} {\bibfield
  {journal} {\bibinfo  {journal} {JCAP}\ }\textbf {\bibinfo {volume}
  {1907}}\bibfield  {number} {\bibinfo  {number} { (07)},\ \bibinfo {pages}
  {012}},\ }\Eprint {https://arxiv.org/abs/1803.03266} {arXiv:1803.03266
  [hep-ph]} \BibitemShut {NoStop}%
\bibitem [{\citenamefont {Abed~Abud}\ \emph {et~al.}(2022)\citenamefont
  {Abed~Abud} \emph {et~al.}}]{DUNE:2022aul}%
  \BibitemOpen
  \bibfield  {author} {\bibinfo {author} {\bibfnamefont {A.}~\bibnamefont
  {Abed~Abud}} \emph {et~al.} (\bibinfo {collaboration} {DUNE}),\ }\bibfield
  {title} {\bibinfo {title} {{Snowmass Neutrino Frontier: DUNE Physics
  Summary}},\ }\href@noop {} {\  (\bibinfo {year} {2022})},\ \Eprint
  {https://arxiv.org/abs/2203.06100} {arXiv:2203.06100 [hep-ex]} \BibitemShut
  {NoStop}%
\end{thebibliography}
\end{document}